\documentstyle[11pt,psfig]{article}
\def\reference{\parskip 0pt\par\noindent\hangindent 0.5 truecm}
\def\s{{\rm\thinspace s}}
\def\km{{\rm\thinspace km}}
\def\kms{\hbox{$\km\s^{-1}\,$}}

\begin{document}
\title{Public Release of 2dF data from the Fornax Cluster Spectroscopic
Survey
}
\author{M.J.\ Drinkwater$^{1}$ \and
  C.\ Engel$^{1}$ \and
  S.\ Phillipps$^{2}$ \and
  J.B.\ Jones$^{3}$ \and M.J.\ Meyer$^{1}$
} 
\date{To appear in the {\em  Anglo-Australian
Observatory Newsletter}}
\maketitle

{\center
$^1$ School of Physics, University of Melbourne, Victoria 3010,
Australia\\[3mm]
$^2$ Department of Physics, University of Bristol, 
Bristol BS8 1TL, UK\\[3mm]
$^3$ School of Physics \& Astronomy,  University of Nottingham,
Nottingham NG7 2RD, UK\\
}

\section{Spectroscopy of Everything}

Almost 20 years ago Morton, Krug \& Tritton (1985) made the first
complete spectroscopic survey of a given region of sky. They surveyed
stellar objects in a 0.3 deg$^2$ region of sky to a limit of $B=20$,
measuring 600 objective prism spectra with the UKST and 100 slit
spectra with the AAT. All the objects measured were normal stars
except for a small number of QSOs and white dwarfs, the latter being
used to put new constraints on the white dwarf luminosity function.

Thanks to the arrival of the 2dF spectrograph on the AAT, we have
recently completed the first stage of a complete spectroscopic survey
more than one order of magnitude larger than the Morton et al.\ study,
measuring 7000 spectra in a $2\pi$ deg$^2$ area as part of our study
of the Fornax Cluster. In this article we describe the public release
of 3600 spectra from our first field. We hope that this public release
will encourage colleagues making surveys for rare objects to choose
these fields, as much of the follow-up spectroscopy that might be
required is available from our data.

Our 2dF Fornax Cluster Spectroscopic Survey (FCSS; see Drinkwater et
al.\ 2000a) was designed to make the most complete census possible of
low-luminosity galaxies in the Fornax Cluster. As well as the 
conventional low-surface brightness dwarfs, we already had evidence
(Drinkwater \& Gregg 1998) that very compact, high surface brightness
dwarf cluster galaxies had been missed in previous work, so we took
the unusual step of observing {\em all} objects in each 2dF field,
both resolved (``galaxies'') and unresolved (``stars''). In this way we
avoided any morphological bias as to what a cluster galaxy should look
like. Observing all the ``stars'' in each 2dF field leads to a large
increase in the number of targets observed: within our magnitude
limits ($16.5<b_J<19.8$) over 50\% of the objects in each 2dF field
are stars. Thanks to the flexibility of 2dF, however, this only leads
to a small increase in total observing time as we schedule the stars
at times when we could not usefully observe galaxies, such as twilight
or at high airmass or through cloud.

\begin{figure}
\hfil \psfig{file=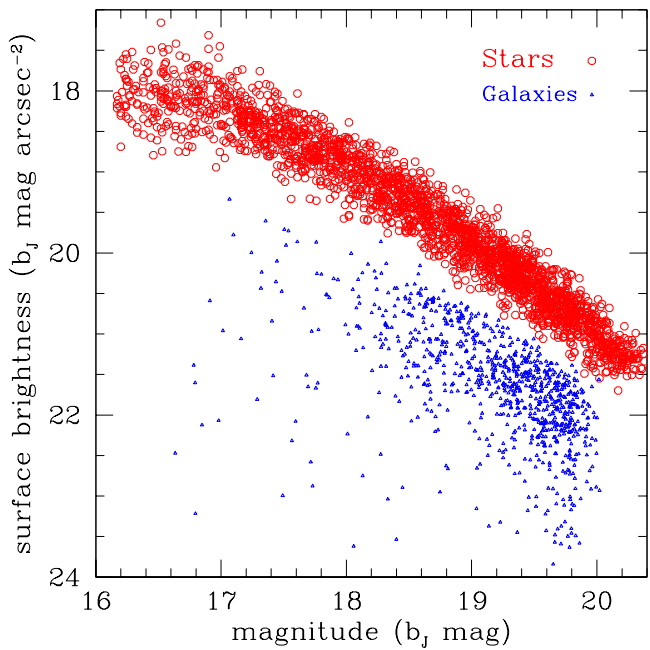,angle=0,width=9cm}
\caption{Surface brightness-magnitude diagram for the objects we have
successfully observed in our first 2dF field. Resolved objects
(``galaxies'') are plotted as (blue) filled triangles and unresolved
objects (``stars'') as (red) open circles. The surface brightness
measurements are based on exponential fits to the photographic data
and are only indicative for the stars (see Drinkwater et al.\ 2000a).
\label{fig-sbm}}
\end{figure}

\section{Status of the Survey}

As of 2001 January we have completed observations of two of our
four planned 2dF fields centred on the Fornax Cluster. We have
analysed the data from our first field and have released
it for public access as described below. The properties of the sample
are summarised in Fig.~1, a plot of surface brightness against
magnitude for all the objects we successfully observed.  The main
all-object sample was selected in the magnitude range
$16.5<b_J<19.8$. We extended our observations of the unresolved
objects (``stars'') to a slightly fainter magnitude limit of
$b_J\approx 20.2.$  There is some incompleteness for galaxies with
surface brightness lower than about 23.5 $b_J$ mag arcsec$^{-2}$ as we
were unable to measure their redshifts (see Drinkwater et al.\ 2000a).

The diversity of objects in our sample is illustrated by Fig.~2, a
cone diagram of the sample in which the distance axis is scaled (as
$z^{1/4}$) so as to display both Galactic stars and the most distant
QSOs. Note that although we avoid the use of conventional
morphological classifications in our analysis (we use redshift,
spectral signatures and luminosity instead), we do show the
morphological classifications in both Fig.~1 and Fig.~2.  The presence
of unresolved objects (``stars'') in both the Fornax Cluster and among
the background field galaxies demonstrates the importance of our
all-object strategy. These objects would have been missed by
conventional galaxy surveys only looking at objects that are resolved
on UKST sky survey plates.

\begin{figure}
\hfil \psfig{file=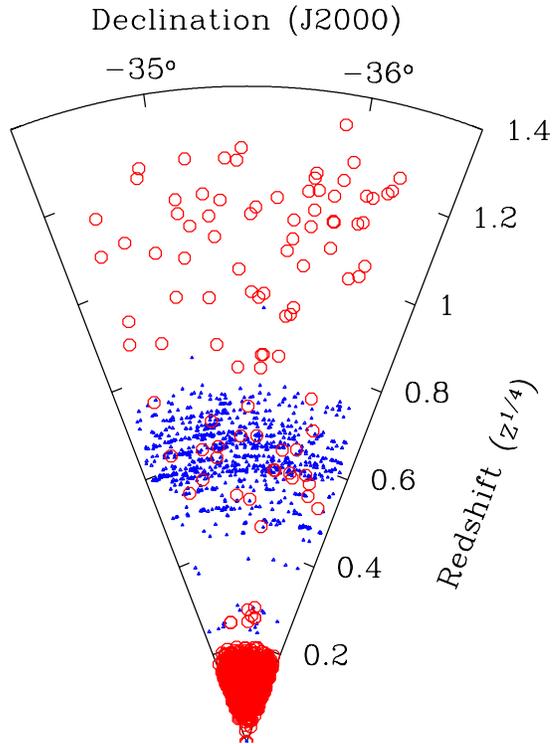,angle=0,width=7cm}
\caption{Cone diagram of the objects we successfully observed in our
first 2dF field. The ``redshift'' axis measures
$z^{1/4}$ in order to display the four orders of magnitude in distance
spanned by our sample.  Resolved objects (``galaxies'') are plotted as
(blue) filled triangles and unresolved objects (``stars'') as (red)
open circles. Note that unresolved sources appear in the Fornax
Cluster ($z^{1/4}\approx0.25$) and among the background field galaxies
($z^{1/4}\approx0.6$).
\label{fig-cone}}
\end{figure}

The scientific highlight of our results to date has been the discovery
of a new population of Fornax Cluster dwarf galaxies so compact they
were previously mistaken for Galactic stars (Drinkwater et al.\
2000b). They can be seen among the more normal, resolved Fornax
Cluster dwarf galaxies in Fig.~2.  These "ultra-compact dwarf" (UCD)
galaxies are unlike any known type of stellar system. They are smaller
and more concentrated than any known dwarf galaxy, but are 2-3
magnitudes more luminous than the largest Galactic globular clusters.
Numerical simulations have shown that the UCDs could have been formed
by tidal stripping of nucleated dwarf galaxies in close orbits around
the central galaxy of the cluster, NGC~1399 (Bekki, Couch \&
Drinkwater 2001). We are making detailed follow-up observations of the
UCDs with {\em Hubble Space Telescope} imaging to measure their sizes
along with high resolution spectroscopy on the VLT to measure their
velocity dispersions and hence masses.

Other results from the early stages of our survey include the first
large QSO sample selected purely from spectroscopic identification
without any pre-selection bias (Fig.~3; see Meyer et al.\ 2001) and
the identification of a population of $L_*$ compact field galaxies
also unresolved from the ground (Drinkwater et al.\ 1999). We have
also shown that some low surface brightness galaxies previously
assumed to be members of the Fornax Cluster are in fact background
objects (Jones et al., in preparation).

\begin{figure}
\hfil \psfig{file=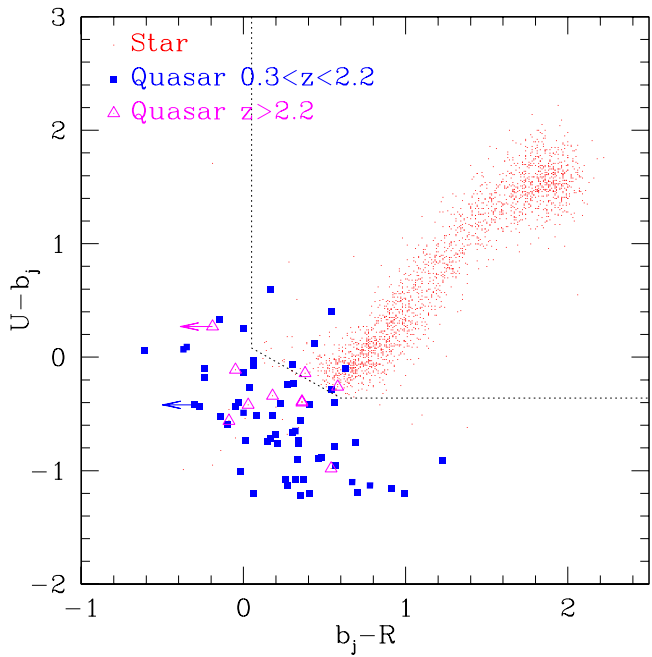,angle=0,width=9cm}
\caption{U$b_j$R colour-colour diagram of QSOs and stars from our
first field. The multicolour selection cutoff used by the 2dF QSO
Redshift Survey is shown by the dotted line.  We derive completeness
values for this multicolour selection of $90 \pm 4$ per cent for
$0.3<z<2.2$ quasars and $80^{+9}_{-13}$ per cent for $z > 2.2$ quasars
(see Meyer et al.\ 2001 for details).  The two quasars in the sample
too faint to appear in the R data are plotted with arrows and a
location corresponding to an upper magnitude limit of R=20.3.
\label{fig-qso}}
\end{figure}

\section{Public Web Access}

The 3600 2dF spectra from our first field are now available for public
access at our web site\\ http://astro.ph.unimelb.edu.au/data/ in
Melbourne. The main function of the web site is to allow searches of
our database by position on the sky and to return our 2dF spectra and
redshift measurements of the selected objects. The complete database
of measurements is available from the website as a plain text file as
well as additional help material and references. The database consists
of all the objects for which we have measured reliable redshifts
(about 95\% of the sample at present). Our 2dF spectra were all
obtained with the same observing configuration as for the 2dF Galaxy
Redshift Survey: a wavelength coverage of 3600--8010\AA\ and a
resolution of 9\AA\ (dispersion of 4.3\AA\ per pixel). The spectra
have typical signal-to-noise ratios of 10 or more per pixel.  The
redshifts were measured by cross-correlation with star and galaxy
templates; this gives velocities accurate to about 65\kms\ for
galaxies (see Drinkwater et al.\ 2000a for details).

The database is searched by entering a co-ordinate and search radius,
with a table listing the parameters of any objects observed within
that radius being returned. The parameters listed include magnitudes
and classifications from the APM Catalogues (Irwin, Maddox \& McMahon,
1994) as well as our own redshift measurements.  Following our
philosophy of avoiding morphological bias in classifications, we do
not specifically classify objects in the database (although the APM
classifications are listed for reference). Instead we recommend the
use of our spectroscopic redshifts (in conjunction with the magnitudes
if necessary) to make classifications on a more physical basis, e.g.\
classifying objects with redshifts less than 700\kms\  as Galactic
stars.

To obtain the 2dF spectrum of any object returned in the output table,
click on the ``plot'' button for that object.  When you click on the
plot button, the spectrum will appear in a new window.  The spectrum
can then be replotted with different axis limits if necessary and
downloaded in text, GIF, PostScript or FITS format. An example of the
PostScript output is shown in Fig.~4.

\begin{figure}
\hfil \psfig{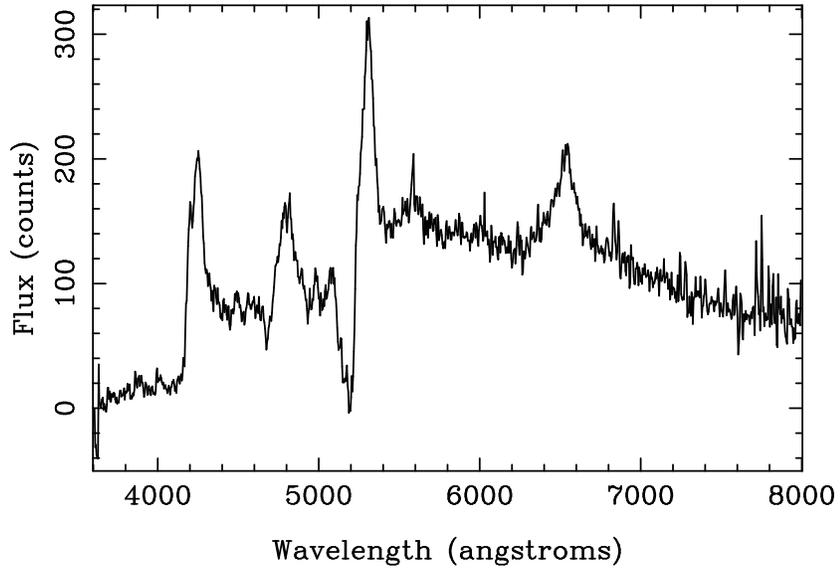}
\caption{A $z=2.43$ QSO from our sample (see Meyer et al.\ 2001), as
plotted by the web interface.
\label{fig-qso67}}
\end{figure}

\section*{References}

\reference Bekki, K., Couch, W.J., Drinkwater M.J., 2001, ApJ, in press
\reference Drinkwater, M.J., Gregg, M.D., 1998, MNRAS, 296, L15
\reference Drinkwater, M.J., et al., 1999, ApJ, 511, L97
\reference Drinkwater, M.J., et al., 2000a, A\&A, 
\reference Drinkwater M.J., Jones, J.B., Gregg, M.D., Phillipps, S.,
2000b, PASA, 17, 227
\reference Irwin M.J., Maddox S., McMahon R., 1994, Spectrum, 2, 14
\reference Meyer M.J., Drinkwater M.J., Phillipps S., Couch W.J., 2001,
MNRAS, in press
\reference Morton D.C., Krug P.A., Tritton K.P., 1985, MNRAS, 212, 325

\end{document}